# Magnetization and magnon excitation energies of the magnetic semiconductors EuTe and EuO on the basis of the renormalized spin wave theory

J. Achleitner

Institute of Semiconductor and Solid State Physics, Johannes Kepler University Linz, Austria

## ABSTRACT

We present analytic expressions for the temperature dependent magnetization and magnon dispersion relation of the antiferromagnetic (AFM) EuTe and the ferromagnetic (FM) EuO. These bulk semiconductors represent concentrated spin systems for which the interaction between magnons has to be taken into account. We do this using the renormalized spin wave theory. A higher order Green's function according to Tjablikov is used for their description. As a result, we obtain a modified Bloch- $T^{3/2}$ law, for low temperatures and high magnetic fields. A full analytic expression is given for the sublattice magnetization of the AFM EuTe ($B_0=0$), with a Néel temperature of $T_N = 9.81$ K. The magnetization curves $\sigma_N(T)$ of EuO (for $0 \leq T \leq 0.68 T_C$) and $\sigma_R(T)$ (for $0.82 T_C < T < T_C$) agree well with experiment. The spin wave excitation energy perfectly matches experimental data from inelastic neutron scattering. In the case of EuTe the magnon excitation energies exhibit a characteristic maximum. For $q = 0$, an energy gap $E_g$ occurs in the spin wave spectrum, a consequence of the AFM exchange interaction $J_2(T)$. Such energy gaps exist in the spectrum of magnon excitation energies only in systems with AFM interactions.

# INTRODUCTION

The Eu-chalcogenides, EuX (with X= O, S, Se, Te, indicating the anions) crystallize in the cubic rock salt structure (NaCl) where the $Eu^{2+}$-ions form a face centered cubic (fcc) structure [1].

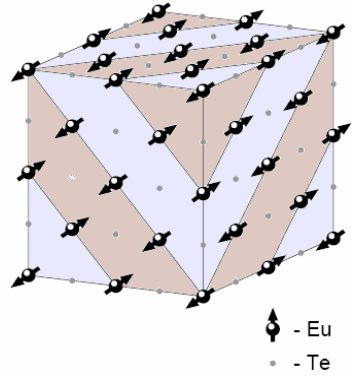

FIG. 1: Chemical and magnetic unit cell of EuTe (fcc) with its AFM type II structure consisting of ferromagnetically ordered (111)-planes and AFM spin sequences along the [111]-directions.

The magnetic properties are determined by the half filled 4f-shells of the magnetic $Eu^{2+}$-ions. Every $Eu^{2+}$-ion shows a magnetic moment of $\mu_B$, where $\mu_B$ stands for the Bohr-magneton. The magnetic 4f moments of EuTe are highly localized at the $Eu^{2+}$ lattice sites (S = 7/2). For EuTe, the exchange interaction of the 4f spins on a given (111)-plane is FM for the nearest neighbours (NN; $J_1>0$, $z_1=12$), AFM for next nearest neighbours (NNN; $J_2<0$, $z_2=6$) and FM for NNNN ($J_3>0$, $z_3=24$), where $J_1$, $J_2$, $J_3$ are the corresponding exchange integrals, and $z_i$ the numbers of neighbours of that type.

The EuX family is characterized by high spin quantum numbers with S = 7/2 and represent concentrated spin systems for which it is necessary to consider the interaction between magnons.

In the literature for the AFM semiconductor EuTe the parameters $J_1$ and $J_2$ for the critical magnetic field never fulfil a necessary $B_c$ with 7.2 T. Examinations of temperature-dependent magnetization in EuX-systems neglect the spin wave interaction and use simple Green's functions in the structure of $G_{ij}(E) = \ll S_i^+ ; S_j^- \gg_E^{ret}$. For the determination of magnetization with $S \geq 1$ it's not possible to use a Green's function according to the expressed relation $G_{ij}(E)$. For the first time, analytic expressions of the magnetization and the magnon excitation energies are given depending on temperature and wavenumber, respectively for the two magnetic semiconductors EuTe (AFM) and EuO (FM) (under the influence of external fields and their exchange parameters). A complete presentation of the sublattice magnetization of the AFM EuTe (bulk system, $B_0$=0 T) is given. To describe the interaction between the magnons. we use the renormalized spin wave theory. A higher Green's function according to Tjablikov is used to describe them. For the AFM semiconductor EuTe the exchange parameters $J_1$ and $J_2$ fulfil a condition for the critical magnetic field $B_c$ with 7.2T (see eq. (35)).

## II. THEORETICAL METHODS

### 1) Application of the theory to concentrated spin- systems

The Eu-chalcogenides are well described as concentrated localized model systems, i.e. each cation site of the crystal lattice is occupied by a localized magnetic moment. We are able thus to describe these systems by the localized Heisenberg - Hamiltonian.

In many antiferromagnets including EuTe, the spins of magnetic atoms strongly interact with each other. In concentrated spin systems, the role of this interaction is essential. Tyablikov's approach[6] and the method of the renormalized spin waves by Dyson [7], [8] consider in their approximations the magnon interactions and can therefore be applied to higher temperatures, too.

So far, the temperature-dependent magnetization process has been determined only numerically and without including the spin wave interactions. Most of these treatments make use of the method of Green's functions in the structure of $G_{ij}(E) = \ll S_i^+; S_j^- \gg_E^{ret}$ [2], [3], [4] ( here $G_{ij}(E)$ is the retarded Green's function for pairs of spins $S_i$ and $S_j$ ).

In this paper, and for the first time, the magnon interaction of both magnetic semiconductors EuTe and EuO are treated using the renormalized spin wave theory.

## 2) Hamiltonian

The seven 4f-electrons of the $Eu^{2+}$-ions in the whole family of the undoped EuX systems build localized magnetic moments $\mathbf{m}_i = \gamma_s \mathbf{S}_i$ , which can be perfectly described by the isotropic 3D-Heisenberg-Hamiltonian ($\mathbf{m}_i$ is the magnetic moment at the lattice site $R_i$ , $\gamma_s$ = gyromagnetic factor).

Anisotropies in the magnetic exchange interactions are characterized on the basis of the dipole-dipole interaction [9]. For single-axis anisotropies (as in the case of EuX [2]) the analysis can be limited to an effective anisotropy field $B_A$

$$H_A = \gamma_s B_A \sum_{i=1}^{N} S_i^z \qquad (1)$$

with $\gamma_s = \mu_B g_j / \hbar$ and $\mu_B = (-e\hbar)/2m_e$ (e = elementary charge, $m_e$ = mass of free electron and $g_j$ (= g) is the Landè factor). The value of the effective $B_A$-field is $B_A = 0.4$ T [9]. With the corresponding Zeeman term, the whole Hamiltonian in the AFM EuTe with regard to both sublattices A and B is formulated as follows:

$$H = -\frac{1}{\hbar^2} \sum_{ij}^{i \neq j} J_{ij} \mathbf{S}_i \cdot \mathbf{S}_j + \gamma_s (B_0 + B_A) \sum_{i \in A} S_i^z + \gamma_s (B_0 - B_A) \sum_{j \in B} S_j^z \qquad (2)$$

The first term includes the FM / AFM coupling between all neighbours, the second and third term characterize the Zeeman- and anisotropy part of the two sublattices, respectively.

### 3) Magnetization and Green function

The central function in magnetic systems is the spontaneous magnetization $M(T, B_0)$:

$$M(T,B_0) = -\frac{1}{V}\frac{\mu_B g_j}{\hbar}\sum_{i=1}^{N}\langle S_i^z \rangle_{T,B_0} \tag{3}$$

which is determined by the expectation value $\langle S_i^z \rangle$ of the z-component of the spin operators $S_i$. (V = sample volume). The thermodynamic expectation value is given by $\langle A \rangle = Sp(\rho A)$ (where $\rho$ is the density operator). For the determination of $\langle S_i^z \rangle(T,B_0)$, we start with the spin correlation function:

$$\langle S_i^+ S_i^- \rangle = \hbar^2 S(S+1) + \hbar \langle S_i^z \rangle - \langle (S_i^z)^2 \rangle , \tag{4}$$

which is valid for all spin quantum numbers ($S \geq 1$) and known as a relation between the spin operators $S_i^\pm$ and $S_i^z$, as well as their spin quantum numbers $m_s$. Here are $S_i^+$, $S_i^-$ ladder-operators, $\langle S_i^z \rangle$ is the thermodynamic expectation value of the z-component of $S_i$, and $\langle S_i^+ S_i^- \rangle$ can be handled via the spectral theorem.

We use a proposition of Tjablikow [10], [11] and formulate an extended theorem of Green's functions with:

$$G_{ij}^{(n)}(E) = \ll S_i^- ; (S_j^z)^n S_j^+ \gg_E^{ret} \tag{5}$$

which is needed for the determination of the magnetization for $S \geq 1$. (Here, n is limited to the values n=0,1,2, ..., 2S-1) [12], [13]

### 3A) Equation of motion / Average magnon occupation - function

The equation of motion in energy representation is formulated for magnons (bosons) based on the n-particle Green's function presented in (5). The poles of this Green function agree with the elementary excitation energies E(**q**) of the spin system. The magnon excitation energies then follow as a solution from the equation of motion:

$$E(\mathbf{q}) = \hbar\omega(\mathbf{q}) = |\mu_B| g_j (B_0 + B_A) - \frac{2}{\hbar}\langle S^z\rangle \left[J_0 - J(\mathbf{q})\right] \quad (= E(\langle\sigma_z\rangle, \mathbf{q})) , \tag{6}$$

with the correlation $\langle S^z\rangle = \hbar/2\langle\sigma^z\rangle$ for every FM orientated sublattice (S≥1). ($J_0$ is the superposition of the exchange integrals for NN and NNN, $J(\mathbf{q})$ is the Fourier transformation of the exchange integral and $\omega(\mathbf{q})$ are the dedicated frequencies). Here the average magnon occupation function $\varphi(S)$ is defined as: [2]

$$\varphi(S) \equiv \frac{1}{N}\sum_{\mathbf{q}} \frac{1}{e^{\beta E(\mathbf{q})} - 1} , \tag{7}$$

where $\beta = 1/kT$, and k is the Boltzmann constant. The thermodynamic expectation value $\langle S^z\rangle$ of the spin operator therefore is: [14]

$$\langle S^z\rangle_{S(\geq 1)} = \hbar\frac{(1+S+\varphi)\varphi^{2S+1} + (S-\varphi)(1+\varphi)^{2S+1}}{(1+\varphi)^{2S+1} - \varphi^{2S+1}} = \langle S_i^z\rangle_{S(\geq 1)} \tag{8}$$

and it is valid for the FM oriented lattice (sublattice) with S ≥ 1. This relation now serves as an additional starting point for the determination of the temperature-dependent magnetization taking into account the interaction between the magnons.

### 4) Interaction between spin-waves

The starting point for the investigation the Eu-chalcogenides is the "renormalized spin wave theory". In this theory the interaction between the spin waves is taken into account by applying the Dyson-Maléev transformation. [15], [16]

By applying a self-consistent Hartree-Fock approximation to describe the interaction term in our Hamiltonian, we get as result an interaction term which becomes temperature dependent via the exchange constants $J_1$ and $J_2$. These functions are interconnected via the temperature-dependent function A(T), $\gamma_q$, as well as the occupation number $n_q$:[17], [18], [19]

$$J_i \rightarrow J_i(T) = J_i(1 - A_i(T)) \tag{9}$$

with

$$A_i(T) = \frac{1}{NS} \sum_{\mathbf{q}_i} (1 - \gamma_{\mathbf{q}}^{(i)}) \langle n_{\mathbf{q}_i} \rangle \tag{10}$$

[20], [21], [22] (J(T) = temperature-dependent exchange function). Here, $\gamma_q^{(i)}$ (i = 1,2) is the geometric structure factor and is defined as:

$$\gamma_{\mathbf{q}}^{(i)} = \frac{1}{z_i} \sum_{\delta_i}^{N_i} e^{i\mathbf{q}\mathbf{r}_{\delta_i}} \tag{11}$$

where $z_1$, $z_2$ are the numbers of nearest and next nearest spin neighbours.

## 5) Analytical representation of magnetization

For the determination of an analytical solution of the temperature dependent magnetization (sublattice magnetization), we have to apply a transformation of the average magnon occupation function φ(S), eq. (7) in the continuum version [5]. Two equations (for S ≥ 1, FM - sublattice) must be solved iteratively:
(i) thermodynamic expectation value of the $S^z$ spin operator, eq. (8); and: (ii) the average magnon occupation-function φ(S), with:

$$\varphi(S) = \frac{1}{N} \frac{V}{(2\pi)^3} \sum_{n=1}^{\infty} e^{-n\beta |\mu_B| g_j B_0} \int d^3\mathbf{q} \, e^{-n\beta \frac{2}{\hbar} (-\langle S^z \rangle) [J_0 - J(\mathbf{q})]} \tag{12}$$

This is the most general starting point for the determination of the magnetization of a spin system with high spin quantum numbers. Here $J_0$ is replaced by $J_0 = J_1 z_1 + J_2 z_2$. This results in a coupled correlation with which the magnetization in magnetic semiconductors can be determined and it can be applied also to Mn-systems (S=5/2), Eu-chalcogenides (S=7/2) etc. J(q) is the Fourier transformation of the exchange function $J(\mathbf{q}) = \frac{1}{N}\sum_{ij} J_{ij} e^{i\mathbf{q}(\mathbf{r}_i - \mathbf{r}_j)}$. For europium-chalcogenides, it is necessary to consider the influence of spins of the NN as well as the NN spin neighbours.

The processing of the summations in eq. (11) is now done for the fcc-lattice structure of EuTe (a is the lattice constant) in the case of the crystallographic $q_{[100]} = (q,0,0)$, $q_{[010]} = (0,q,0)$ and $q_{[001]} = (0,0,q)$ directions. So we get for the difference:

$$J_0 - J(\mathbf{q}) = 8J_1 + 2J_2 - 8J_1 \cos(\frac{qa}{2}) - 2J_2 \cos(qa) \qquad (13)$$

### III. RESULTS

#### 1) Spin-waves and interaction terms

In the interaction case, the spin wave stiffness D becomes temperature-dependent[2] in accordance with $D(T) = D \cdot (1-A(T))$. For low temperatures $A(T) \to 0$ and thus we can start for $T < 10$ K with $D(T) \approx D$.

Based on the renormalized spin wave theory the "coupling function" $A_i(T)$ results in eq. (10). Because of the exponential function in the integrand we are able to extend the integration with respect to the wave number over the whole q-space to infinity. Introducing Riemann's $\zeta$-function $\zeta(\alpha)$ we obtain:

$$A_1(T) = \frac{Va^2}{NSz_1(\sqrt{\pi})^3}\left[\frac{3}{16}\zeta(\frac{5}{2})(\frac{kT}{D})^{\frac{5}{2}} - \frac{5}{512}a^2\zeta(\frac{7}{2})(\frac{kT}{D})^{\frac{7}{2}} + \frac{7}{24576}a^4\zeta(\frac{9}{2})(\frac{kT}{D})^{\frac{9}{2}}\right]. \qquad (14)$$

The interaction is characterized by the power terms of temperature like $T^{5/2}$, $T^{7/2}$ and $T^{9/2}$. The determination of $A_2(T)$ (next nearest neighbours, $z_2=6$) follows from the same procedure. Summing up all terms, the formula for the temperature-dependent part $A_2(T)$ becomes:

$$A_2(T) = \frac{Va^2}{NSz_2(\sqrt{\pi})^3} \left( \frac{3}{16}\zeta(\frac{5}{2})(\frac{kT}{D})^{\frac{5}{2}} - \frac{5}{128}a^2\zeta(\frac{7}{2})(\frac{kT}{D})^{\frac{7}{2}} + \frac{7}{1536}a^4\zeta(\frac{9}{2})(\frac{kT}{D})^{\frac{9}{2}} \right) \quad (15)$$

In line with the determination of $A_1(T)$ and $A_2(T)$ (or $J_i(T)$, respectively), this means that $J_0 - J(\mathbf{q})$ is completely determined as well.

## 2) $I^{st}$ case: Small magnon energies, *i.e.* φ(S) << 1

For very low temperatures (T~ 0), the magnon occupation function $\varphi(S) \ll 1$ and thus $\varphi(S) \ll S$. In order to derive the thermodynamic expectation value of the spin operator $S^z$ (eq. (8)), we use an expansion with respect to the parameter $\varphi = \varphi(S)$. This results in:

$$\langle S^z \rangle_{S(\geq 1)} = -\hbar S + \hbar \varphi(S) + O(\varphi^2) \quad (16)$$

(for the temperature region $T \in [0, \frac{3}{4}T_N]$ ).

Thus, for the average magnon occupation function, the integral in eq. (12) must be determined. With the eqs. (9), (10) and (11) we do the integrations arising in φ(S) through the 1$^{st}$ Brillouin zone. The evaluation provides in the case of vanishing external magnetic fields ($B_0 = 0$) an analytic expression:

$$\varphi(S) = \Lambda_1 \frac{1}{|\langle\sigma^z\rangle|^{\frac{3}{2}}} T^{\frac{3}{2}} + \Lambda_2 \frac{1}{|\langle\sigma^z\rangle|^{\frac{5}{2}}} T^{\frac{5}{2}} + \Lambda_3 \frac{1}{|\langle\sigma^z\rangle|^{\frac{7}{2}}} T^{\frac{7}{2}} \quad (17)$$

with the leading temperature terms $T^{3/2}$, $T^{5/2}$ and $T^{7/2}$. This provides the complete determination of the sublattice magnetization for the 1st temperature region (T = 0....7.9 K) in accordance with the previous relation and gives

$$\sigma_N(T) = \frac{1}{M_{max}} M(T, B_0) = S - \frac{1}{|f(\Lambda_1)|^{\frac{3}{2}}} \Lambda_1 T^{\frac{3}{2}} - \frac{1}{|f(\Lambda_1)|^{\frac{5}{2}}} \Lambda_2 T^{\frac{5}{2}} - \frac{1}{|f(\Lambda_1)|^{\frac{7}{2}}} \Lambda_3 T^{\frac{7}{2}} \qquad (18)$$

with:

$$f(\Lambda_1) = 2S - \frac{1}{\sqrt{2}S^{\frac{3}{2}}} \Lambda_1 T^{\frac{3}{2}} \qquad (19)$$

and the parameters $z_1 = 12$, $z_2 = 6$, $S = 7/2$, $a = 6.598 \times 10^{-10}$ m, $k = 1.3806568 \times 10^{-23}$ J/K, $V = 1$ m$^3$, $N = 2.47 \times 10^{28}$. The spin wave stiffness is given by:

$$D = (z_1 J_1 + z_2 J_2) S a^2 \qquad (20)$$

The functions $\Lambda_i$ are also complex temperature dependent terms in eq. (18), (19) and are results of the spin wave interactions.

## 2A)  Modified Bloch law

Bloch's $T^{3/2}$ law, is a well defined starting point for experimental investigations of the temperature dependent magnetization. However, it has the disadvantage that it applies only for diminishing magnetic fields. A correction has been done at first by Dyson[15] who considered further temperature terms containing $T^{5/2}$ and $T^{7/2}$

$$M_B = \frac{m}{S}\left[S - \zeta(\frac{3}{2})\theta^{\frac{3}{2}} - \frac{3}{4}\pi v \zeta(\frac{5}{2})\theta^{\frac{5}{2}} - \omega \pi^2 v^2 \zeta(\frac{7}{2})\theta^{\frac{7}{2}} + O(\theta^{\frac{9}{2}})\right]. \qquad (21)$$

Here $M_B$ is the spontaneous magnetization, m is the magnetic moment of each atom, S = spin of each atom, $\zeta(\alpha)$ = Riemann's zeta-function, $\theta$ = a dimensionless temperature (= $T/2\pi T_c$), $\omega$ is a numerical coefficient (for the three types of cubic lattices), ν is a geometrical factor for the simple, face-centered and body-centered cubic lattices. Here the temperature terms $T^{3/2}$, $T^{5/2}$ and $T^{7/2}$ show only constant coefficients (proportional to the Riemann zeta-functions). $M_B$ is valid only up to half of the Néel / Curie temperature, a field-free system ($B_0 = 0$), for the special cases with $S = 1/2$, $S = 1$ and $S = \infty$ and no spin wave interaction is taken into account.

If only low temperatures are relevant, the terms with $T^{5/2}$ and $T^{7/2}$ can be neglected in (18), so we obtain a considerably better, modified Bloch $T^{3/2}$ law that takes the interaction of the spin waves in the low temperature region into account and in addition remains valid even for high magnetic fields.

$$M(T,B_0) = \frac{N}{V}\frac{\mu_B g_j}{2}\left(2S - \frac{1}{\sqrt{2}S^{\frac{3}{2}}}\Lambda_1 T^{\frac{3}{2}}\right). \tag{22}$$

Eq. (22) is generally valid for an AFM in fcc-structure. A more comprehensive investigation concerning the influence or increase of the spinwave-interaction shows a significant reduction in the transition temperature. Fig. 2 provides the whole temperature characteristics for case (I) in accordance with (17) and φ(S) << 1.

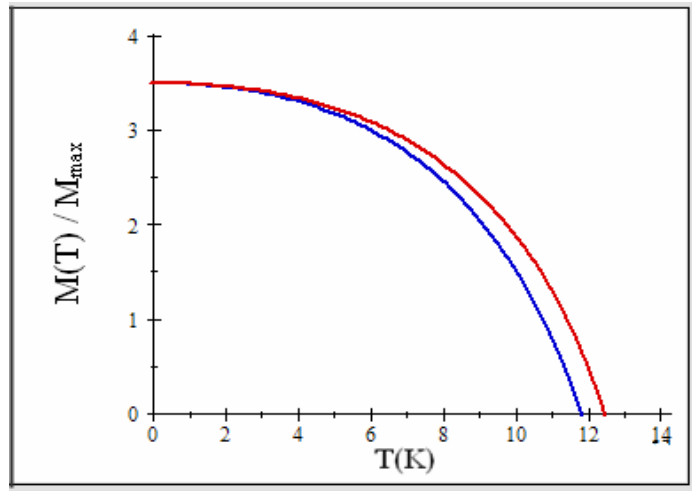

FIG. 2: Distribution of magnetization for case (I) of low temperatures (magnon occupation function with φ(S)

≪1): graphical representation of the modified Bloch $T^{3/2}$ – law according to eq. (22). Red curve: without spin wave interaction, blue curve: with interaction. (The latter reduces also the Néel-temperature)

The modified $T^{3/2}$ (Bloch) law was derived in a form that removes these limitations, thus making it also applicable for $B_0 \neq 0$ and spin wave interactions.

### 3) II$^{nd}$ case: High magnon energies, *i.e.* φ(S) ≫ 1

Near the transition temperature (T ≤ $T_N$), φ(S) according to eq. (12) becomes large, φ(S) ≫ 1. So we get for the expansion of the thermodynamic expectation value of $S^z$ (case of EuX) with respect to 1/φ up to third order:

$$\langle S^z \rangle_{S(\geq 1)} = -\hbar \frac{S(S+1)}{3\varphi(S)} \qquad (23)$$

Here the maximum of magnons are excited. For the continuum representation of φ(S) (field-free case, $\beta E(q) \ll 1$) we obtain the average magnon occupation function φ(S) as:

$$\varphi(S) = -\frac{kT}{\langle \sigma^z \rangle} F_{-1} - \frac{1}{12} \frac{1}{kT} \langle \sigma^z \rangle F_1 + \frac{1}{720} \frac{1}{k^3 T^3} \langle \sigma^z \rangle^3 F_3 \quad , \qquad (24)$$

where the three $F_i$-functions represent the integrations over the exchange functions with:

$$F_{-1} = \frac{1}{(2\pi)^3} \frac{V}{N} \int_{1.BZ} d\mathbf{q} \frac{1}{J_0 - J(\mathbf{q})} \quad , \qquad F_1 = \frac{1}{(2\pi)^3} \frac{V}{N} \int_{1.BZ} d\mathbf{q} [J_0 - J(\mathbf{q})]$$

$$F_3 = \frac{1}{(2\pi)^3} \frac{V}{N} \int_{1.BZ} d\mathbf{q} [J_0 - J(\mathbf{q})]^3 \qquad (25)$$

The respective integrations extend in each case over the 1$^{st}$ Brillouin zone.

### 3A) Solutions of the spin equation

If $\varphi(S)$ is known, we can solve eq. (16). This leads to

$$\frac{3}{2}\langle\sigma^z\rangle = \frac{S(S+1)}{\frac{kT}{\langle\sigma^z\rangle}F_{-1} + \frac{1}{12}\frac{1}{kT}\langle\sigma^z\rangle F_1 - \frac{1}{720}\frac{1}{k^3T^3}\langle\sigma^z\rangle^3 F_3} \quad (26)$$

This algebraic equation provides 4 solutions, but only 2 of them have physical relevance. The result is a solution for $\langle\sigma^z\rangle$ in the region of transition temperature ($T \lesssim T_N$):

$$\langle\sigma^z\rangle = \sqrt{30k^2\frac{T^2}{F_3}F_1 - 6\sqrt{25k^4\frac{T^4}{F_3^2}F_1^2 - 20k^3\frac{T^3}{F_3}\left(\frac{2}{3}S(S+1) - kTF_{-1}\right)}} \quad (27)$$

We denote $\langle\sigma^z\rangle$ in the following chapters as $\sigma_R(T)$. $\sigma_N(T)$ (eq. (18)) and $\sigma_R(T)$ are the dimensionless magnetizations ($\langle S^z\rangle = \hbar/2\langle\sigma^z\rangle$) in the case of $\varphi(S) \ll 1$, or $\varphi(S) \gg 1$, respectively.

By doing the integrations in eq. (25) for $F_{-1}(T)$, $F_1(T)$ and $F_3(T)$ - introducing spherical coordinates - $\sigma_N(T)$ and $\sigma_R(T)$ are fully given. The $F_m(T)$-functions characterize clearly the interactions of the spin waves based on the renormalized spin wave theory.

### 4) Magnon dispersion relation - EuO

The presence of a significant anisotropy as observed in the case of EuTe (with $B_A = 0.4T$) appears to be of minor importance for EuO according to present knowledge. Spin wave excitations show a significant influence of the nearest and next nearest neighbours on the coupling mechanism (FM / AFM) [24], [25], [26].

Expressed by the exchange functions according to the renormalized spin wave theory, eqs. (9) and (10), the magnon excitation-energy E(**q**) for any temperature T=τ is given in meV by ($B_0=0$)

$$E_\Omega(\mathbf{q},T) = \langle \sigma^z \rangle_{T=\tau} \left( \frac{[d_1]_{T=\tau}}{1 \times 10^{-23}} (1 - \frac{1}{z_1}(4 + 8\cos(\frac{qa}{2}))) + \frac{[d_2]_{T=\tau}}{1 \times 10^{-23}} (1 - \frac{1}{z_2}(4 + 2\cos(qa))) \right) 6.2415 \times 10^{-2} \quad (28)$$

where the coupling functions $d_1(T,J_1)$, $d_2(T,J_2)$ (of NN and NNN) are given by

$$d_1 = z_1 J_1 (1 - A_1(T)) \quad , \quad d_2 = z_2 J_2 (1 - A_2(T)) \quad (29)$$

The relevant value of the magnetization $\langle \sigma_z \rangle$ is needed in the magnon dispersion relation eq. (28).

### 5) Magnon excitation energies of the Heisenberg AFM – EuTe

For the determination of the magnon dispersion relation of an AFM, two interacting sublattices with opposite spin direction must be analysed. [29], [30], [31], [32], [33]

Starting point is the full Hamiltonian (formulated for a short-range interaction) with an additional Zeeman- and anisotropy term and a localization of the magnetic 4f-moments of the EuTe in the $Eu^{2+}$-lattice sites.

The Heisenberg-operator (2) is set in the presentation for NN (FM-coupling) and NNN (AFM-coupling)

$$H = -\frac{J_1}{\hbar^2} \sum_{ij}^{i \neq j} \mathbf{S}_i . \mathbf{S}_j + \frac{|J_2|}{\hbar^2} \sum_{ik}^{i \neq k} \mathbf{S}_i . \mathbf{S}_k .$$

It's not possible to solve our Heisenberg model exactly for the general case. This is the reason why a transformation is used – in accordance with Holstein-Primakoff – in the form of:

$$S_i^+ = \hbar\sqrt{2S}\varphi(n_i)a_i \quad , \quad S_i^- = \hbar\sqrt{2S}a_i^+\varphi(n_i) \quad , \quad S_i^z = \hbar(S - n_i) . \quad (30)$$

where $n_i = a_i^+ a_i$ denotes the occupation number operator, with the operator function $\varphi(n_i)$:

$$\varphi(n_i) = \sqrt{1 - \frac{n_i}{2S}} \ . \tag{31}$$

In this equation, the operator function is treated in accordance with the linear spin wave theory. The Hamiltonian built by virtue of this transformation is represented by the wave number k for the respective sublattices A and B after Fourier transformation. So we take a transformation that additionally is canonical and linear, in accordance with Bogoliubov and find the dispersion relation for EuTe:

$$\begin{aligned} E_\alpha^+(\mathbf{k}) = \hbar\omega_\alpha(\mathbf{k}) &= \frac{1}{2}(b_A - b_B - a(\mathbf{k}) + 2SJ_1 z_1) + \\ &+ \frac{1}{2}\sqrt{(b_A + b_B - a(\mathbf{k}) - 2c(\mathbf{k}) + 2SJ_1 z_1)(b_A + b_B - a(\mathbf{k}) + 2c(\mathbf{k}) + 2SJ_1 z_1)} \end{aligned} \tag{32}$$

with:

$$\begin{aligned} a(\mathbf{k}) &= 2SgJ_1(\mathbf{k}) \qquad \text{FM}-\text{interaction} \\ c(\mathbf{k}) &= -2SgJ_2(\mathbf{k}) \qquad \text{AFM}-\text{interaction} \end{aligned} \tag{33}$$

and the abbreviations:

$$b_A \equiv 2z_2 |J_2| S - \gamma_s \hbar(B_0 + B_A) \quad , \quad b_B \equiv 2z_2 |J_2| S + \gamma_s \hbar(B_0 - B_A) \ . \tag{34}$$

Here $b_A$, $b_B$ are field indicators in the case of magnon excitation energies of AFM EuTe. The spin wave interaction in eqs. (32), (33), (34) is taken into account (based on the renormalized spin wave theory of the Dyson-Maleév transformation) in accordance with the temperature-dependence of the exchange parameters $J_1$ and $J_2$.

### IV. COMPARISON with EXPERIMENT

#### 1) Analytical representation of the sublattice magnetization

With the relations $\sigma_N(T)$ and $\sigma_R(T)$, we want to determine in a next step a graphical representation of the sublattice magnetization of the AFM EuTe (bulk system) for $B_0 = 0$. Including all the given parameters, we obtain the temperature-dependency of σ.

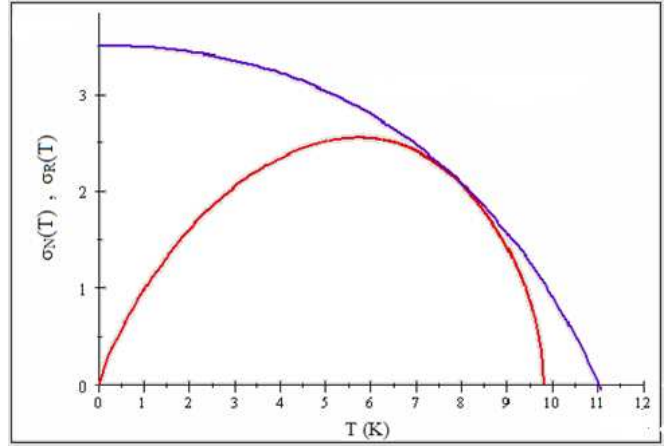

FIG. 3: Magnetization in the case (I) and (II) for $\varphi(S) \gg 1$ (red curve: $\sigma_R(T)$). The blue curve, $\sigma_N(T)$, represents the situation $\varphi(S) \ll 1$ with the maximum $\sigma_N(T=0) = 7/2$. Tangential intersection of both functions in the neighbourhood of $T = 7.9$ K.

The numerically determined solution of the equation $\langle \sigma_z \rangle = \sigma_R(T) = 0$ leads to a Néel-temperature of $T_N = 9.813$ K, with the values of the exchange constants. The value then experimentally determined provides a temperature point at which the magnetization vanishes – of $T_N = 9.81$ K. The maximum occurs at $T = 0$ K with $\langle \sigma_z \rangle = 7/2$.

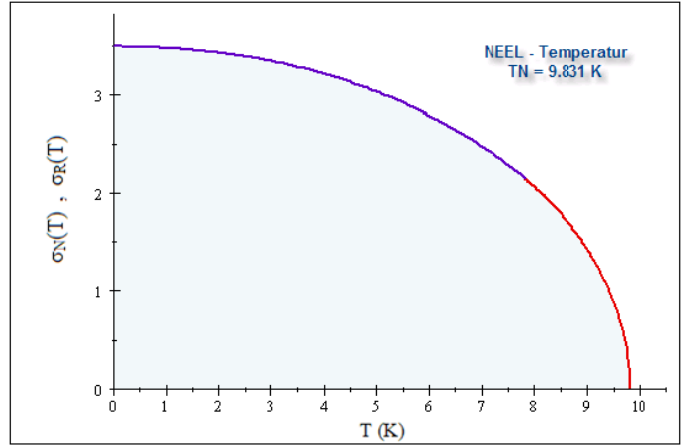

FIG. 4: Full temperature dependent sublattice magnetization of the AFM EuTe, bulk system. Analytical solution according to eqs. (18) and (27) for an average magnon occupation function $\varphi(S) \ll 1$ (blue curve, valid for T $\in$ [0,7.9 K]), $\varphi(S) \gg 1$ (red curve, up to $T_N = 9.813$ K)

For the exchange constants $J_1$ and $J_2$ the modified values of Masset and Callaway[23] are used. They have to fulfil the relation

$$|B_c| = \frac{4m_s}{\mu_B g_j}(z_1 J_1 + z_2 J_2) \tag{35}$$

for the critical magnetic field of $B_c = 7.2$ T for EuTe (based on the double sum of the Heisenberg operator in (2)), *i. e.*, the magnetic field, for which all spins are completely oriented in a parallel arrangement. For both parameters $J_1$ and $J_2$ of EuTe, we have:

$$J_1/k = 0.07 K \quad , \quad J_2/k = -0.25515 K \ ,$$

and thus: $J_1 = 9.6646 \times 10^{-25}$ J , $J_2 = -3.5228 \times 10^{-24}$ J.

### 2) Magnetization and magnon dispersion relation for Europium-oxide (EuO)

Based on the theory developed for EuTe, we aim to verify the statements that have been made so far for EuO. Experiments based on inelastic neutron scattering have shown that EuO as well as EuS represent ideal Heisenberg ferromagnets. The relevant exchange constants $J_1$ and $J_2$ were determined by inelastic neutron scattering by Dietrich and Passell [27], [28] to be:

$$J_1/k = 0.606 K \quad , \quad J_2/k = 0.119 K .$$

EuO has a Curie temperature of $T_C = 69.3$ K.

The derived analytical solutions of the temperature-dependent magnetization are the basis for both magnetization curves, $\sigma_N(T)$ and $\sigma_R(T)$, for $B_0=0$. For this we find two temperature regions:

$$\begin{array}{cc} \text{Region(I) for } \sigma_N(T) & \text{Region(II) for } \sigma_R(T) \\ (\text{"Low-temp"}) \ 0 \leq T \leq 0.68 T_C & (\text{"High-temp"}) \ 0.82 T_C \leq T \leq T_C \\ \varphi(S) \ll 1 & \varphi(S) \gg 1 \end{array} \quad (36)$$

The sub-functions of $\sigma_N(T)$ and $\sigma_R(T)$ are just as presented before under eqs. (18), (19), (20), (29) and (14), (15).

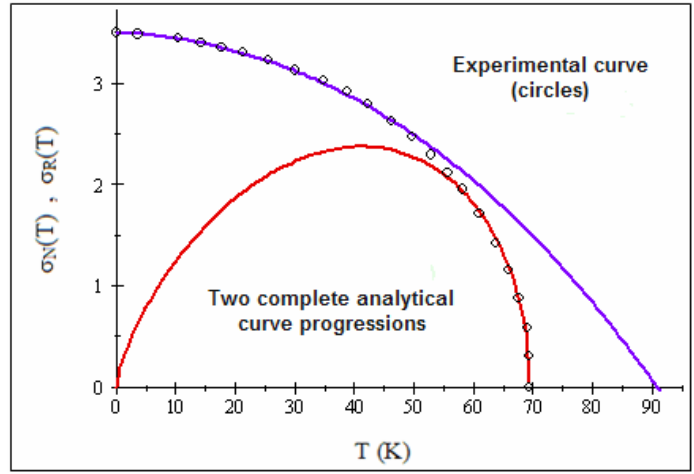

FIG. 5: Temperature-dependent magnetization for EuO; experimental data (circles)[36] in comparison to both analytical solutions. Curve structure of the region (I),

$\sigma_N(T)$ – presented as blue curve. Magnetization structure

of region (II), $\sigma_R(T)$ – red curve.

The circles in the graph represent the experimental magnetization behaviour. Theoretical data for the temperature-dependent magnetization are given by the respective exact structures of $\sigma_N(T)$ and $\sigma_R(T)$.

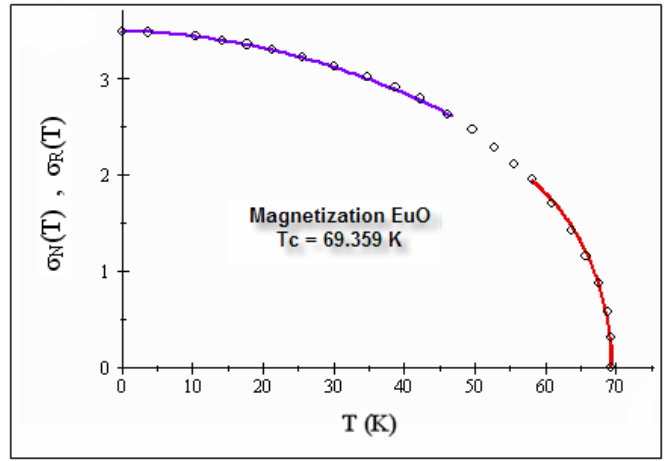

FIG. 6: Full analytical structure of the temperature-dependent magnetization of EuO, for $\varphi(S) \ll 1$ (left curve) and $\varphi(S) \gg 1$ (right curve). Comparison of the two analytical solutions ($\sigma_N(T)$, $\sigma_R(T)$) with the experimental structure (circles). $\sigma_R(T) = 0$ yields the Curie temperature $T = T_C$.

We can see perfect agreement of experimental data with $\sigma_N(T)$ from 0 K up to a temperature of 47 K and with $\sigma_R(T)$ for the region from 57 up to 69.36 K.

The magnetization analytically determined for both cases $\varphi(S) \ll 1$ und $\varphi(S) \gg 1$ shows for EuO too an excellent agreement in the corresponding temperature regions on the basis of the theory presented here. The Curie temperature obtained for this purpose is $T_C = 69.359$ K and follows numerically from $\sigma_R(T) = 0$.

## 2A) Dispersion relation EuO

The only energies for EuO in the experimental determination of the magnon excitation energy was performed by Passel [27] by means of the inelastic neutron scattering for T=5.5 K. The spin wave propagation was experimentally examined for $q_{[100]}$, $q_{[110]}$ and $q_{[111]}$. For a given wave number a magnetization value of $<\sigma_z> = \sigma_N$ (T=5.5 K) = 3.4920 follows from eq. (28). The circles drawn in Fig. 7 were taken from ref. 27.

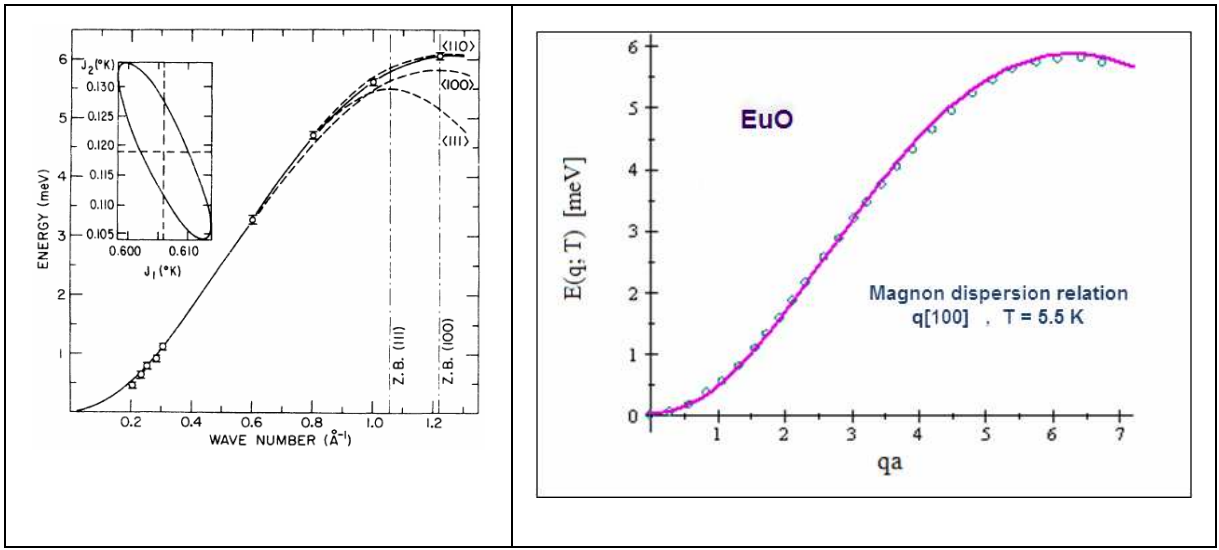

FIG. 7: Full analytical structure of the magnon dispersion relation of EuO (wave number $q_{[100]}$), T=5.5 K. Comparison of the analytical solution (full curve) with the experimental data (on the left) derived from inelastic neutron scattering (circles).

The functional representation shows for EuO a perfect agreement with the experimental results. The positions of the maximum are consequently identical for a wave number with $q_m = 1.2222 (Å)^{-1}$ in the theoretical case and with $q_m = 1.22 (Å)^{-1}$ for the experimental situation.

## 2B) Magnon excitation energy depending on the temperature

So far, the energy of the magnons has been examined by the coupling functions $d_1$ and $d_2$ for a fixed temperature $T = \tau$ ($\tau$ plays the role of a fixed parameter). In order to obtain a dependence in the form $E=E_\Omega(T)$, we examine $E_\Omega(T;q)$ for a fixed wave number q. ($E_\Omega(T;q)$ = magnon excitation energy of EuO). In comparison with the experiments we use E(q) for q = 0.2(Å)$^{-1}$. For the magnon dispersion relation the complete magnetization $<\sigma_z>$ is necessary for the representation of the whole FM region, eq. (28). In this situation, the joint of region (I) ($\sigma_N(T)$) with region (II) ($\sigma_R(T)$) lies at T = 54 K. The "open temperature region" between $\sigma_N(T)$ and $\sigma_R(T)$ (see fig. 6) is represented by a variation of the exchange parameters $J_1$ and $J_2$ in respect to a correspondingly adapted function $\sigma^V_R(T)$.

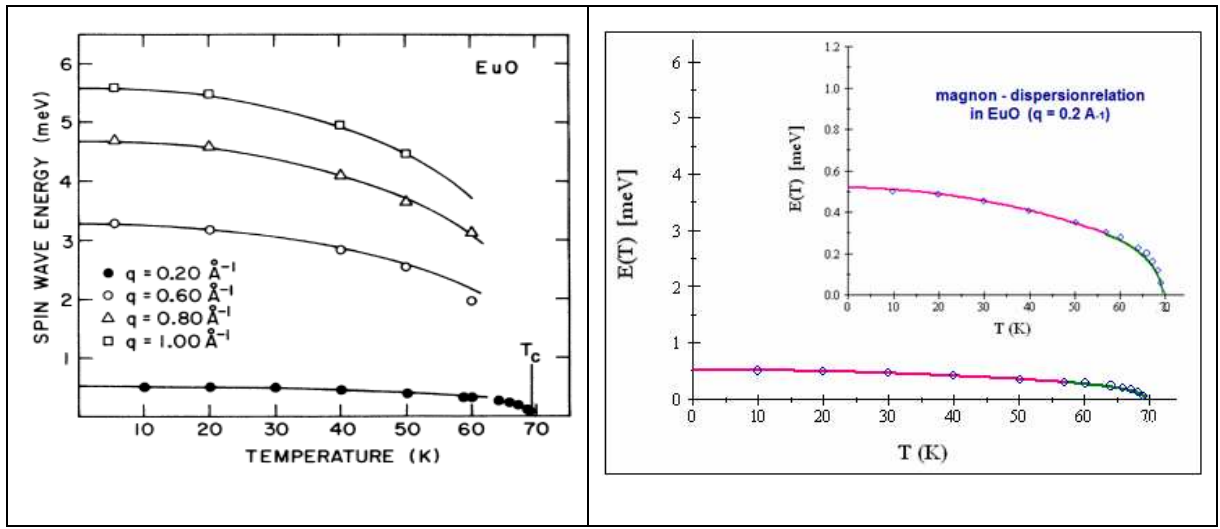

FIG. 8: Comparison of the theoretical dependence E(T;q) with the experimentally determined data of the magnon dispersion relation. The graphical comparison of the magnon energies of EuO is given for a wave number q=0.2 (Å)$^{-1}$

The temperature-dependent representation of the magnon energy shows a perfect agreement of the theory with the experimental data.

### 3) EuTe – Experimental relation, variation of temperature and energy gap

So far, no experimental measurement of the magnon dispersion relation has been reported for the antiferromagnet EuTe. However, it shares in its functional behaviour in the sense that there are some qualitative similarities with the experimentally determined excitation curve of iron-oxide $Fe_{1-x}O$ ( = wustite), which is measured at temperatures of liquid helium by the inelastic neutron scattering. $Fe_{1-x}O$ [34] shows indeed equivalent properties as EuTe concerning its essential characteristics (antiferromagnetism, NaCl-structure etc.).

The dispersion curves experimentally determined with inelastic neutron scattering (measurement bars) are now compared (fully marked lines in fig. 9) with the theory used in this case (a phenomenological Heisenberg spin wave model). The prominent direction is the [100]-direction.

The important rise from the position of the energy gap in the magnon energy is, like for iron-oxide $Fe_{1-x}O$, a consequence of both types of interactions. We consider the first Brillouin zone of the fcc lattice of EuTe, especially along the Γ-X -axis with a distance Δ. The borders of this zone result from the claim $δ = q_m a = 2π$ and shows for $E(q;T)$ a fully symmetrical behaviour in the interval with $δ \in [-2π, 2π]$.

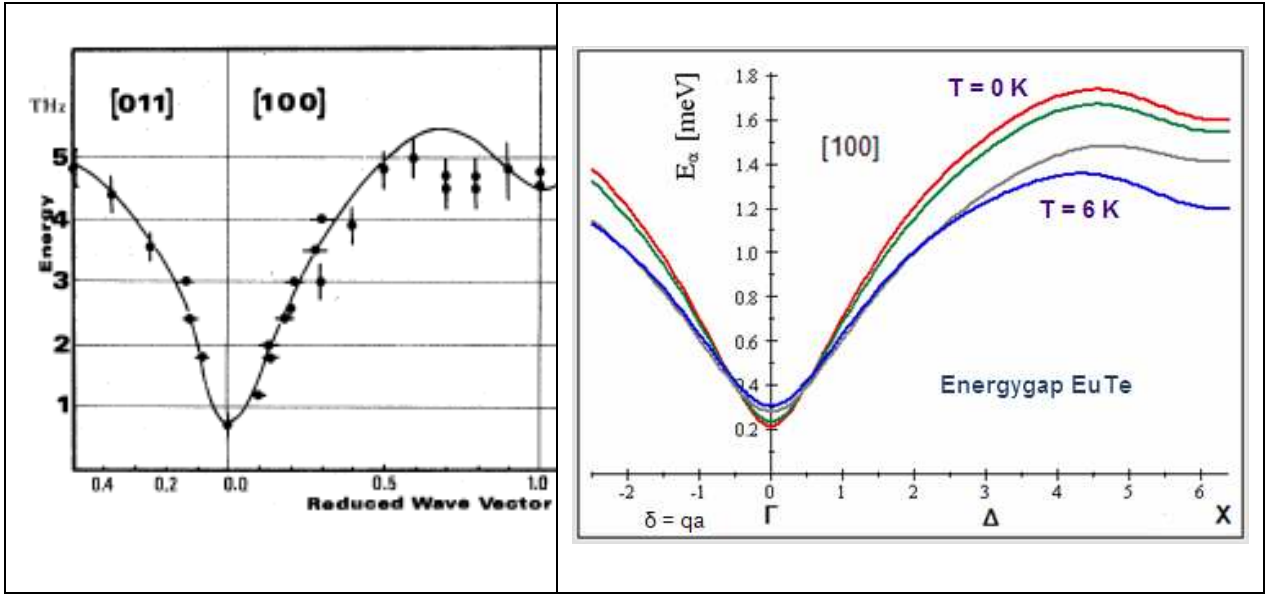

FIG. 9: Magnon dispersion relation of EuTe for the wave numbers $q_{[100]}$, $q_{[010]}$, $q_{[001]}$ in the full physically relevant structure in comparison with the magnon excitation energy of the iron-oxide $Fe_{1-x}O$ (= wustite) (representation on the left). The graphic presents the behaviour of the magnon dispersion relation with variation for 4 different temperatures. T=0 K (red curve), T=2 K (green curve), T=4 K (grey curve), T=6 K (blue curve). For T = 0 K the smallest energy gap is obtained. With increasing temperature (T = 6 K), the gap increases accordingly. A local minimum appears at the border of the 1. BZ (X-point) in $q_{[100]}$ direction.

In the next step, we investigate the system for different temperatures at $B_0 = 0$ T. The magnon excitation energies for EuTe (eq. 32) are plotted for 4 different temperatures (T= 0 K, 2 K, 4 K and 6 K). The variable is $\delta = qa$. The dispersion structure for variable temperatures shows the diagram presented in Fig. 9. As in the case of wustite we find here too a prominent minimum at the position $qa=2\pi$.

For q = 0, an energy gap $E_g$ typical for the AFM EuTe is caused in the spin wave spectrum. This energy gap is defined at the position of the minimum spin wave energy. We evaluate this gap from eq. (32) for $E_\alpha(q;T)$, where the dependence on the wave vector **q** is given by both functions a(**q**) and c(**q**) (see eq. 33). We obtain thus the following relation for the energy gap $E_g$ in the spin wave spectrum:

$$E_g = \hbar\omega(\mathbf{q}=\mathbf{0},T)_{B_0=0} = \sqrt{\left(\mu_B g_j B_A\right)^2 - 24S\mu_B g_j B_A |J_2(T)|} \qquad (37)$$

For our calculations we take the Landé factor as $g_j = 2$, the anisotropy field with $B_A = 0.4$ T and for the exchange value (NNN) $J_2 / k = -0.25515$ K, according to sec. IV.1). The results show a significant dependence of $E_g$ on $B_A$. This relation reveals a characteristic temperature at which the energy gap disappears. The energy gap has an essential influence on the physical properties like specific heat and internal energy.

### 3A)   EuTe – Variation of the magnetic field

We investigate $E_\alpha(q; B_0)$ in the case of 4 magnetic fields (Fig. 10) and a vanishing system temperature (T = 0K). The external magnetic field provokes a parallel shift of the energy branch and a significant increase of the magnon energy.

It must be mentioned, that for relatively small fields, the minimum for $\delta = 2\pi$ – due to the AFM influence – becomes flatter and flatter. (The external field $B_o$ will align spins with opposite orientation). For a characteristic value of $B_0$ the small "depression" disappears completely. For EuTe this starts from about $B_0 = 0.5$T.

Comparison with the AFM Praseodymium (Pr):

(1) For EuTe, there is a steady and monotonous rise of the magnon excitation energy in the region $\delta = \pi/2$ up to $\delta = 3\pi/2$.

(2) Similarly to the case of EuTe, this behaviour appears also in the excitation spectrum of antiferromagnetic Praseodymium ($Pr^{3+}$) and Europium ($Eu^{3+}$) – when a large crystal field is added. [35]

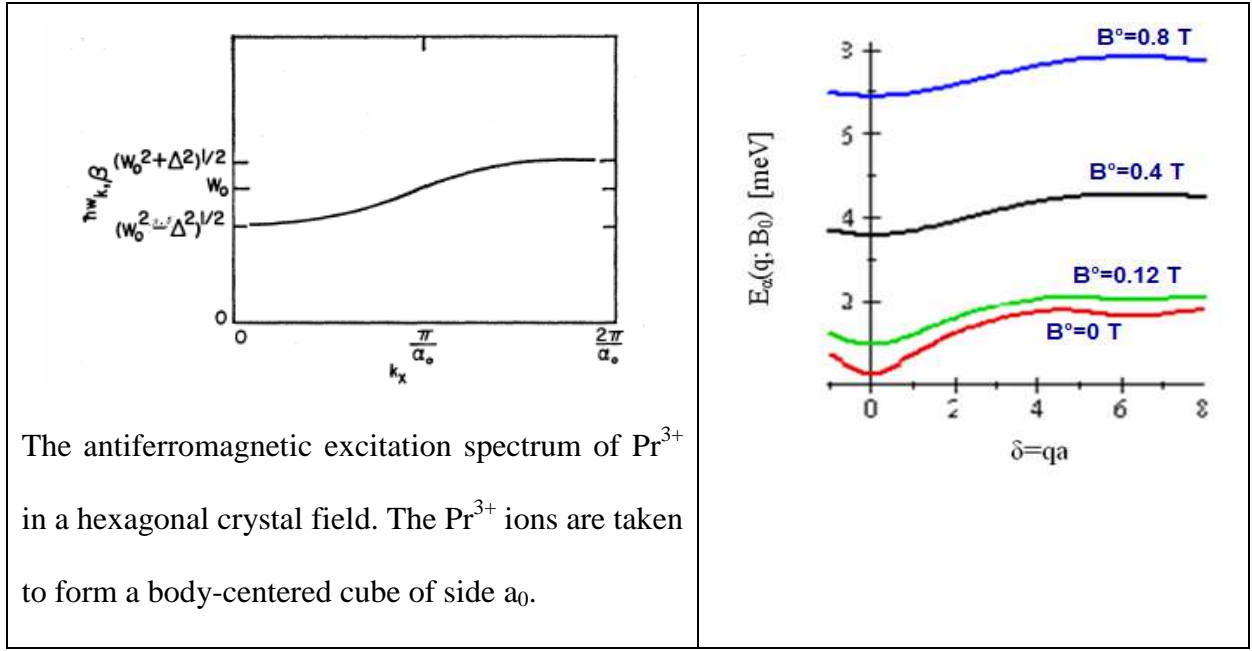

The antiferromagnetic excitation spectrum of $Pr^{3+}$ in a hexagonal crystal field. The $Pr^{3+}$ ions are taken to form a body-centered cube of side $a_0$.

FIG. 10: Excitation spectrum of antiferromagnetic $Pr^{3+}$ (left) in comparison with that calculated for EuTe (T=0K) for 4 different magnetic fields (right): $B_0 = 0$ T (red curve), $B_0 = 0.12$ T (green curve), $B_0 = 0.4$ T (black curve), $B_0 = 0.8$ T (blue curve)

### 3B)  EuTe - Variation of $J_1$ and $J_2$

We examine now the bulk system EuTe at the maximal values of magnetization (T ∈ [0,2K]) for vanishing magnetic fields ($B_0$=0T). The behavior of the magnon excitation energies are studied for the case of a purely FM and AFM condition compared to the situation of a combined FM / AFM interaction. We represent the characteristics of the magnon excitation energies with a given magnetic structure in the three cases

$$J_1 \neq 0 \quad J_2 = 0 \quad \text{\textit{FM interaction}}$$
$$J_1 = 0 \quad J_2 \neq 0 \quad \text{\textit{AFM interaction}}$$
$$J_1 \neq 0 \quad J_2 \neq 0 \quad \text{\textit{complete interaction}}$$

As a result, the magnon dispersion relation shows for these types of interactions the response shown in Fig. 11.

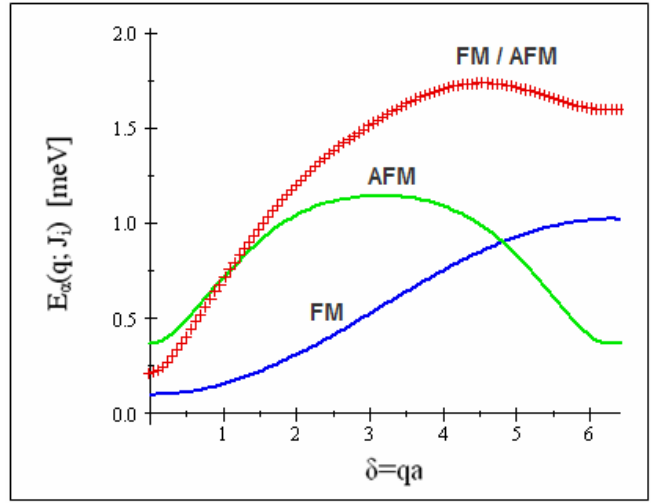

FIG. 11: Effects of possible exchange-interactions on the magnon excitation energies. FM-interaction (blue curve), AFM (green curve), situation for $J_1$ and $J_2 \neq 0$ (red curve)

The dispersion branch resulting from the FM-interaction shows its maximal value for $\delta = qa = 2\pi$. It is exactly this point where the experimental curve, in respect to the AFM-interaction shows a local minimum. For $\delta \in [0, 9/2]$, the overlap shows a steady increase of both energies as a consequence of the FM and AFM interactions.

The subsequent occurrence of a minimum is a result of the relatively steep drop of the magnon energy (for $\delta > 9/2$) due to the AFM spin structure.

## V. CONCLUSIONS / SUMMARY

The magnetic semiconductors EuX are characterized by high spin quantum numbers with S = 7/2. A higher Green's function according to Tjablikov is used to describe them.

This work treats the two magnetic semiconductors EuTe (AFM) and EuO (FM). These semiconductors represent concentrated spin systems for which it is necessary to consider the interaction between the magnons. Our process is treated by the renormalized spin wave theory.

For the first time, analytic expressions are derived for the magnetization and magnon excitation energies as a function of temperature (under the influence of external fields and their exchange parameters).

A complete presentation of the sublattice magnetization of the AFM EuTe (bulk system, $B_0=0$ T) is given. The resulting Néel-temperature with $T_N$ = 9.813 K is in perfect agreement with the experimental value of $T_N$ = 9.81 K.

The analytical solution determined for the magnetization can also serve to derive a modified Bloch $T^{3/2}$ law, which is now valid for low temperatures, high magnetic fields and the interaction between the spin waves is taken into account.

Of all Eu-chalcogenides (EuX) EuO shows the highest phase transition temperature (a Curie temperature $T_C$=69.36 K). Both magnetization curves $\sigma_N(T)$ and $\sigma_R(T)$ gives an accurate agreement with the experiment in the considered temperature region: for $\sigma_N(T)$ from 0 K up to a temperature of 47 K (0 $\leq$T $\leq$0.68 $T_C$ ); those of $\sigma_R(T)$ apply to the interval of 57 to 69.36 K ( 0.82 $T_C$ $\leq$T$\leq$ $T_C$).

The spin wave excitation energy shows a perfect agreement with the values derived from inelastic neutron scattering – and this in their wave number dependence, as well as in their temperature dependence.

This gives a perfect confirmation of the derived results based on the analysis of the used Green function by Tjablikov. The results can be used for the FM EuS too by taking the allocated system parameters like $J_1$ and $J_2$, a, N and so on.

In the AFM EuTe, a characteristic maximum is reflected in the magnon excitation energies with a subsequent drop to qa=2π. An external magnetic field (T=0 K) triggers a parallel shift of the energy branch and an increase in the magnon energy.

The resulting "AFM energy depression" disappears when a characteristic field is reached. For q=0, there is an energy gap $E_g$ (which is typical for the AFM EuTe) in the spin wave spectrum. It represents a significant dependence on the anisotropy field $B_A$ and the AFM interaction function $J_2(T)$. The derived relation gives an explanation for the absence of an energy gap in systems with only FM interactions.

## Acknowledgments

I am specially grateful to Eberhard Heiner and Wolfgang Jantsch for interesting discussions and stimulating comments. I also thank Günther Bauer, Wolfgang Heiss and Reinhard Folk for helpful comments.